\documentclass{iopart}

\usepackage{iopams,setstack}
\usepackage{graphicx}
\usepackage{cite}
\usepackage{enumerate}
\usepackage{hyperref}

\newcommand{\bd}{\begin{displaymath}}
\newcommand{\ed}{\end{displaymath}}
\newcommand{\be}{\begin{equation}}
\newcommand{\ee}{\end{equation}}
\newcommand{\ba}{\begin{eqnarray}}
\newcommand{\ea}{\end{eqnarray}}

\begin{document}

\title{A simple projective setup to study optical cloaking in the classroom}

\author{I. Marcos-Mu\~noz and A. S. Sanz}

\address{Department of Optics, Faculty of Physical Sciences\\
Universidad Complutense de Madrid\\
Pza.\ Ciencias 1, Ciudad Universitaria E-28040 Madrid, Spain}

\ead{\mailto{a.s.sanz@fis.ucm.es}}

\begin{abstract}
Optical cloaking consists in hiding from sight an object by properly deviating the light that comes
from it.
An optical cloaking device (OCD) is an artifact that hides the object and, at the same time, its
presence is not (or should not be) noticeable for the observer, who will have the impression of
being looking through it.
At the level of paraxial geometrical optics, suitable for undergraduate courses, simple OCDs can be
built by combining a series of lenses.
With this motivation, here we present an analysis of a simple projective OCD arrangement.
First, a simple theoretical account in terms of the transfer matrix method is provided,
and then the outcomes from a series of teaching experiments carried out with this device,
easy to conduct in the classroom, are discussed.
In particular, the performance of such an OCD is investigated by determining the effect of the
hidden object, role here played by the opaque zone of an iris-type diaphragm, on the
projected image of an illuminated transparent slide (test object).
That is, cloaking is analyzed in terms of the optimal position and opening diameter of a
diaphragm that still warrants an almost unaffected projected image.
Because the lenses are not high-quality ones, the OCD is not aberration-free, which is
advantageously considered to determine acceptable cloaking conditions (i.e., the
tolerance of the device).
\end{abstract}







\section{Introduction}
\label{sec1}

In {\it Harry Potter and the Deathly Hallows} \cite{rowling:HarryPotter:2007},
J.~K.~Rowling tells about the origin of the famous piece of cloaking tissue that
allows Harry Potter, the renowned young wizard apprentice, to hide from unwanted company.
This is just an example taken from modern literature of how the idea of optical cloaking and
invisibility, the capability to hide objects from sight (people in this case), has permeated popular culture
in recent times, although there are many more examples.
For instance, as is well known, a number of optical illusions in fashion during the XIXth and XXth centuries were
based on the reflection properties of mirrors and window glasses \cite{steinmeyer-bk,zepf:physteach:2004}.
It is perhaps because of this capability to both fascinate and amaze at the same time, that the scientific
community has also shown much interest for the cloaking phenomenon and its possible applications
\cite{azanna:optica:2018}, from the stealth technology (widely used in the military
industry in aircrafts or warships, for example) to the design and fabrication of metamaterials
\cite{engheta:PRE:2005,engheta:PRE:2006,smith:Science:2006,shalaev:NatPhot:2007}.

In general, (optical) cloaking conditions are related to a particular geometrical design of the
object to be hidden or the cloaking device that is going to be used to hide the object.
Conversely, such a design strongly depends on the level at which cloaking is required.
This can be done either at the highly sophisticated level of electromagnetic optics
or at the more elementary level of geometrical optics.
At the level of electromagnetic optics, this implies fabricating specific arrangements
that produce a direct, local influence on the phase of the incident light.
This is the basic principle behind the possibility to produce cloaking metamaterials, of much
interest at present, although out of the scope here.
Nevertheless, just to provide a quick glimpse on the issue, it is worth mentioning that
in these new materials cloaking emerges from the collective behavior of repetitive
patterns of assemblies of tiny units, smaller than the typical wavelengths they intend to
influence, instead of the material itself that such units are made of.
By means of a specific design of such assemblies, the electric and magnetic properties
of the material can be controlled, achieving narrow electromagnetic spectral bands,
infinite phase velocities, or negative refractive indexes.
At a pedagogical level, there have been some proposals to introduce the physics of
metamaterials in simple terms with the aid of inexpensive experimental setups, based
on elements available in any undergraduate teaching laboratory and the use of
microwaves \cite{marques:AJP:2011,fleming:AJP:2017} or sound waves \cite{gennaro:AJP:2016}.
Other authors have approached the issue, particularly cloaking properties, from a more
theoretical perspective, through easy-to-implement numerical tools \cite{thompson:AJP:2008} or with
just some elementary theory \cite{longhi:AJP:2017}, also appropriate at the undergraduate level.

At the level of geometrical optics, on the other hand, the cloaking effect can
be achieved in a far simpler fashion, involving either multiple reflections \cite{choi:ApplOpt:2014,howell:video}
or the imaging properties of lens systems \cite{choi:OptExp:2014,choi:OptExp:2015,howell:video2}.
In either case, the working principle is the same: the light coming from an object is redirected in a way
that it reaches the observer's eyes without being affected by the presence of an interposed element,
which remains unnoticeable for the observer.
This is the case, for instance, of the optical cloaking device (OCD) proposed in Ref.~\cite{choi:OptExp:2014},
which is based on the very basic contracting/expanding property of a two-lens system.
More specifically, the waist of an incident light beam coming from the object is reduced
by a telescope-like two-lens configuration and, then, after the light has traveled a certain
distance, the inverted version of the same two-lens configuration expands again the beam waist
to its initial size, which is eventually what the observer is looking at.
In the middle part of the OCD, between the reducer and the expander two-lens systems, the narrowness
of the beam allows to accommodate around it whatever object we wish to hide, because its presence
will not be noticed by the observer.
A nice video demonstration of the experiment can be seen in \cite{howell:video2}.
The same experiment can also be performed with standard lenses, available in any teaching
optics laboratory.
Analogous results can also be obtained with cheaper versions, such as the one specially prepared
in our teaching laboratory for a popular science TV show \cite{orbita-laika}.

Because of its simplicity, it is clear that for an optics instructor the aforementioned
experiment is very appealing at an elementary teaching level, as a way to introduce the concept
of optical cloaking in the classroom in simple terms: just a simple arrangement of a few lenses
and basic theory illustrates very nicely an application of the principles of geometrical optics
in paraxial form, beyond other conventional examples that have been used in optics
laboratories for decades to demonstrate the functioning of telescopes, microscopes,
photo-cameras, the eye, or other optical instruments.
For undergraduate students, on the other hand, these experiments would constitute a beneficial
first-approach to the phenomenon, since it is introduced at a very basic level, without getting into
very sophisticated mathematics or the use of specific optics software.
Now, the question here is how to design an experiment in a way that can be quantitatively meaningful,
is easy to be described and analyzed theoretically, and, very important, does not require expensive
(professional) laboratory material, which is what one usually has at hand in typical teaching
laboratories.
The setup proposed in \cite{choi:OptExp:2014,choi:OptExp:2015,howell:video2} produces
very spectacular results (just what everyone is expecting from this kind of experiences), but our
experience in the laboratory with this kind of OCD is that students lose track very quickly of the
physics behind it, because eventually all is based on apparatent sizes, semi-qualitative analyses
and, of course, the funny cloaking effect when they interpose objects (or even their hands and
faces).
Hence, instead, we considered an observer-independent alternative setup, more appropriate
to conduct quantitative measurements and to study the physics involved (and less prone to
distractions), where cloaking is not directly observed by looking through, but is analyzed in terms
of the effects the hidden object makes on the image of the observed object.

Here, we report on theory and measurements performed with the above mentioned OCD.
The work combines a preliminary theoretical
analysis with a subsequent experimental development of a simple OCD
The theoretical analysis is based on the transfer matrix method applied to optics \cite{pedrotti-bk},
which allows the student to understand the physics of the optical cloaking phenomenon by
investigating imaging through a lens system in a relatively simple fashion (no high knowledge
on the issue is actually required, but just simple matrix algebra).
This analysis allows to determine optimal cloaking conditions.
Accordingly, the simpler, optimal configuration consists of two sets of lenses with different
focal lengths arranged in the form of two Kepler-type telescopes faced by their eyepieces.
The analysis also provides the distance that should be considered between those ``eyepieces''.
With these data, the device is then mounted with cheap resources (as we have confirmed,
there is no need for high-quality, expensive lenses to observe the phenomenon) available in
any teaching laboratory.
In this regard, we have noticed that, even with the lowest-quality lenses available in our laboratory
(formerly used to study the effects of aberrations), the amount of cloaking obtained is already
remarkable.
Furthermore, the presence of spherical and chromatic aberrations is taken advantageously
to determine unambiguously the limits of our optimal cloaking conditions (beyond them,
aberration effects start dominating very quickly the image observed).

The work has been organized as follows.
To be self-contained, in Sec.~\ref{sec2} the basic theory on the transfer matrix method is
introduced, which is based on Gaussian paraxial optics (small-angle approximation) and
therefore does not require expert-level knowledge.
An analysis and discussion of the application of this approach to systems with increasing
number of lenses is also presented in order to settle down the theory that the experiment is
based on.
This approach is applied to the analysis of cloaking with systems with increasing number of
lenses.
The experimental setup considered here is described in Sec.~\ref{sec3}, discussing some of
the main aspects that have been taken into consideration for its implementation as well as
the main outputs.
To conclude, a series of final remarks are summarized in Sec.~\ref{sec4}.


\section{Theoretical analysis}
\label{sec2}

The OCD implemented in Ref.~\cite{choi:OptExp:2014} worked in a very simple manner:
when looking through the OCD, any object observed behind it should be seen as if there
was nothing between such an object and our eyes, even if we introduce another object
inside the OCD.
This behavior can be summarized in terms of the following two conditions:
\begin{enumerate}
 \item The image of the observed object must be direct, virtual, and with unitary
 magnification.

 \item The OCD has some extension, henceforth denoted with $L$, so that the hidden object
 can be accommodated somewhere inside it.
\end{enumerate}
Accordingly, the rays leaving the object will reach our eyes with a minor influence
from the hidden object or the OCD itself.

The variant here proposed works in a similar way, although instead of a virtual image of a
background object, we are going to focus on the real image produced by an illuminated object.
That is, instead of analyzing cloaking by direct observation, we are going to analyze it by
observing a projected image, although the theoretical analysis is equally suitable to both.
Consequently, for optimal cloaking conditions, such an image should be unaffected by the
presence of the OCD itself or an object hidden inside it.

It should also be mentioned that the above condition (i) is not essential regarding
cloaking, unless the OCD is also required to be hidden from sight, i.e., we do not wish
to notice the presence of the OCD, but only the background as it appears when there is
nothing in front of it (along the direct line of sight).
This is worth mentioning, because it enables the possibility to construct alternative
OCDs in a way that, even if their configuration is not exactly the same as the one
reported in \cite{choi:OptExp:2014}, still the cloaking phenomenon can be observed.


\subsection{Elementary aspects of the transfer matrix method}
\label{sec21}

In Gaussian paraxial optics, when dealing with simple lens systems, imaging is typically determined
by means of ray tracing.
An efficient way to tackle the issue when the number of optical elements increases (which, from a
practical point of view, essentially means considering more than two or three lenses) is by making
use of the so-called transfer matrix method \cite{pedrotti-bk}.
This is an easy-to-handle input/output method based on the linear relationship between the object
(input) and its conjugate image (output) independently of the number of optical elements (lenses
and mirrors) accommodated between both.
Such a relationship is given in terms of the so-called $ABCD$ matrix,
\be
 \mathbb{M} =
 \left( \begin{array}{cc}
  A & B \\ C & D
  \end{array} \right) ,
 \label{eqA}
\ee
where each element is directly related with a property of the optical system itself, if the matrix
is computed between its two boundary surfaces, or the imaging process, if it is defined from the
object plane to the image one.

To better understand this basic concept, consider an object point $P_O$ and its conjugate image
point $P_I$.
The point $P_O$ is at a height $h_O$ off the optical axis and $P_I$ is at $h_I$.
Both points can be joined by a swarm of rays, all employing the same time in going from one to the
other, according to Fermat's least time principle.
Let us consider one of such rays.
This ray leaves $P_O$ at an angle $\alpha_O$ with respect to the direction of the optical axis,
and reaches $P_I$ with an angle $\alpha_I$ (also with respect to the optical axis).
Although it is not shown here (but it is not difficult to prove either), $h$ and $\alpha$ are
the only two parameters we need to characterized the imaging process in paraxial optics.
The relationship between the input (object) properties, $(h_O,\alpha_O)$, and the output (image)
ones, $(h_I,\alpha_I)$, is described by a linear matrix transformation, $\mathbb{M}$, which
transfers the former to the latter.
If these properties are recast in vector form, we have
\be
 {\bf p}_I = \left( \begin{array}{c} h_I \\ \alpha_I \end{array} \right)
  = \left( \begin{array}{cc} A & B \\ C & D \end{array} \right)
 \left( \begin{array}{c} h_O \\ \alpha_O \end{array} \right)
  = \mathbb{M} {\bf p}_O .
 \label{eqC}
\ee

According to Eq.~(\ref{eqC}), the height and inclination of the image point are given,
respectively, by
\ba
 h_I & = & A h_O + B \alpha_O ,
 \label{eqE} \\
 \alpha_I & = & C h_O + D \alpha_O ,
 \label{eqF}
\ea
from which it is readily seen that $A$ and $D$ are dimensionless parameters,
while $B$ and $D$ have length and inverse-length dimensions, respectively.
The dimensionality of these matrix elements can easily be understood by noting that
$A$ is related to the linear magnification of the image with respect to the object
(perpendicularly measured from the optical axis, i.e., the ratio $h_I/h_O$), while $D$
is related to the angular magnification, which describes the apparent size with respect
to the object (i.e., $\alpha_I/\alpha_O$).
Regarding the elements $C$ and $B$, they are associated with the positions of the first
and second focal planes of the optical system, $h_I/\alpha_O$ and $h_O/\alpha_I$,
respectively, taking its input and output planes as a reference.


\subsection{Transfer matrix for an $N$-lens system}
\label{sec22}

In the particular case we are going to deal with here, only two types of matrices are needed, namely a matrix
describing the passage of light through a single thin lens, which in paraxial form reads as
\be
 \mathbb{L} =
 \left( \begin{array}{cc}
  1 & 0 \\ - 1/f & 1
  \end{array} \right) ,
 \label{eq1}
\ee
with $f$ being the lens effective focal length, and a translation matrix,
\be
 \mathbb{T} =
 \left( \begin{array}{cc}
  1 & d \\ 0 & 1
  \end{array} \right) ,
 \label{eq2}
\ee
accounting for the transit of ray bundles through an empty space of length $d$
(this can be the space between two consecutive lenses, or just the distance between
the object and the first lens and the distance from the last lens to the image, as will
be seen below).

From the above considerations, (i) and (ii), an ideal OCD should behave analogously to a single
translation matrix with $d = L$, i.e.,
\ba
 h_I & = & h_O + L \alpha_O ,
 \label{eqEocd} \\
 \alpha_I & = & \alpha_O ,
 \label{eqFocd}
\ea
so that the image has the same size and orientation as the object when looked from any
direction (that is, any $\alpha_O \neq 0$).
Thus, let us consider a system of $N$ lenses with their centers aligned along the system
optical axis.
In this system, $f_n$ is the effective focal length of the $n$th lens and $d_{n-1}$ denotes
the distance between the $n$th and $(n-1)$th lenses, with $d_0 \equiv 0$.
Imaging in this system is determined by the matrix product
\be
 \mathbb{M}_N = \mathbb{L}_N \mathbb{T}_{N-1} \mathbb{L}_{N-1}
  \cdots \mathbb{L}_1 \mathbb{T}_2 \mathbb{L}_1
  = \overleftarrow{\Pi}_{n=1}^N \mathbb{S}_n ,
 \label{eq3}
\ee
with
\be
 \mathbb{S}_n \equiv \mathbb{L}_n \mathbb{T}_{n-1} ,
 \label{eq4}
\ee
and where the arrow over the product symbol ($\Pi_n$) denotes that each new
product element $n$ has to be added to the left instead of to the right.
Notice that for $n=1$, we have $\mathbb{T}_0 = \mathbb{I}$, since $d_0 = 0$.

When the explicit form of the matrices (\ref{eq1}) and (\ref{eq2}) is
substituted into (\ref{eq4}), with $f_n$ and $d_{n-1}$ instead of $f$ and $L$,
respectively, the product matrix (\ref{eq3}) reads as
\be
 \mathbb{M}_N =
  \overleftarrow{\Pi}_{n=1}^N
  \left( \begin{array}{cc}
  1 & d_{n-1} \\ - 1/f_n & 1 - d_{n-1}/f_n
  \end{array} \right) ,
 \label{eq5}
\ee
which is of the form
\be
 \mathbb{M}_N =
 \left( \begin{array}{cc}
  A_N & B_N \\ C_N & D_N
  \end{array} \right) .
 \label{eq6}
\ee
This matrix is to be compared with the total translation matrix that represents
the ideal OCD, i.e.,
\be
 \mathbb{M}_N =
 \left( \begin{array}{cc}
  1 & L \\ 0 & 1
  \end{array} \right) ,
 \label{eq7}
\ee
with $L = \sum_{n=1}^N d_{n-1}$.
By comparing matrices (\ref{eq6}) and (\ref{eq7}) element by element, we obtain the
set of equations
\ba
 A_N & = & 1 ,
 \label{eq8a}
 \\
 B_N & = & L ,
 \label{eq8b}
 \\
 C_N & = & 0 ,
 \label{eq8c}
 \\
 D_N & = & 1 ,
 \label{eq8d}
\ea
which are used to design the OCD.
In compliance with Eqs.~(\ref{eqEocd}) and (\ref{eqFocd}),
The fact that $A_N$ and $D_N$ are both unitary means that the image
produced by the device of an object located to its right must also have
unitary lateral and angular magnification (image equal to object) even if
there is a cloaked object inside it.
That is, the picture collected by lens 1 is directly transferred to lens $N$,
without any further optical operation, as it is inferred from the fact that
$B_N = L$.
Moreover, like a telescope, the system is afocal, since $C_N = 0$.

Next we are going to consider systems with $N$ ranging from 2 to 4 in order
to better understand how the cloaking property works.
The case $N=1$ is not considered, because it is the trivial one given by (\ref{eq1})
when instead of a lens we have a very thin plate (with negligible thickness).
Obviously, this can never be an OCD, because there is no room to hide an object
inside it.
We thus need to take into account that the conditions specified by Eqs.~(\ref{eq8a})
to (\ref{eq8d}) are necessary for cloaking, but not sufficient.


\subsection{Two-lens system}
\label{sec23}

For a two-lens system, after proceeding with the product of matrices, the conditions
given by Eqs.~(\ref{eq8a}) to (\ref{eq8d}) read as
\ba
 A_2 & = & 1 - \frac{d_1}{f_1} ,
 \label{eq9a}
 \\
 B_2 & = & d_1 ,
 \label{eq9b}
 \\
 C_2 & = & - \frac{1}{f_1} - \frac{1}{f_2} + \frac{d_1}{f_1 f_2} ,
 \label{eq9c}
 \\
 D_2 & = & 1 - \frac{d_1}{f_2} .
 \label{eq9d}
\ea
This system is analogous to a thick lens with thickness $t = d_1$ and refractive
index $n_L = 1$.
The term $-C_2$ provides us with the equivalent power of the system, also known as Gullstrand's
equation \cite{milton-bk}, from which the system focal length is readily obtained: $f = -1/C_2$.

If we apply the condition (\ref{eq8c}) to the Eq.~(\ref{eq9c}), we
find
\be
 f_1 + f_2 = d_1 ,
 \label{eq10}
\ee
i.e., for the system to be afocal, the distance between the two lenses must be equal to
the sum of their respective focal lengths.
This is precisely the condition that makes afocal a telescope.
However, contrarily to a telescope, where we are interested in large magnification factors,
in the case of the OCD we are looking for a unitary magnification.
Thus, in order to ensure that the magnification elements (\ref{eq9a}) and (\ref{eq9d}) are
nearly unitary, both focal lengths, $f_1$ and $f_2$, must be much larger than $d_1$, so that
$d_1/f_1 \ll 1$ and $d_1/f_2 \ll 1$.
When this condition is satisfied, we find that, although there is room to place an object
inside this two-lens device, the situation is again analogous to the above single-lens
case: cloaking conditions lead to a non-cloaking device formed by two plane-parallel
plates.
Therefore, this is another example where conditions (\ref{eq8a}) to (\ref{eq8d}) are
necessary for cloaking, but not sufficient.


\subsection{Three-lens system}
\label{sec24}

In the case of three lenses, the elements of the matrix $\mathbb{M}_3$ describing
the system read as
\ba
 A_3 & = & 1 - \frac{L}{f_1} - \left( 1 - \frac{d_1}{f_1} \right) \frac{d_2}{f_2} ,
 \label{eq11a}
 \\
 B_3 & = & L - \frac{d_1 d_2}{f_2} ,
 \label{eq11b}
 \\
 C_3 & = & - \frac{1}{f_1} - \frac{1}{f_2} - \frac{1}{f_3}
           + \left( \frac{d_1}{f_1} + \frac{d_2}{f_3} \right) \frac{1}{f_2}
 \nonumber \\ & &
           + \left( L - \frac{d_1 d_2}{f_2} \right) \frac{1}{f_1 f_3} ,
 \label{eq11c}
 \\
 D_3 & = & 1 - \frac{L}{f_3} - \left( 1 - \frac{d_2}{f_3} \right) \frac{d_1}{f_2} ,
 \label{eq11d}
\ea
with $L = d_1 + d_2$.
As before, we find that $f_2$ must be very large in order that the condition (\ref{eq8b})
satisfies, which means that the second lens essentially behaves as a thin glass layer that
does not affect all the other components of the system.
Actually, this leads to a two-lens system analogous to the previous one, for which the
same equations hold after replacing $f_2$ in Eqs.~(\ref{eq9a})--(\ref{eq9d}) by $f_3$.

There is, however, an alternative non-trivial solution.
By inspecting the magnification terms, $A_3$ and $D_3$, we notice that they display
some symmetry when $f_1$ and $f_3$ are exchanged, and also when the same is done with
$d_1$ and $d_2$.
If we apply conditions (\ref{eq8a}) and (\ref{eq8d}), we obtain the following relation
\be
 \frac{f_1}{f_3} = \frac{d_1}{d_2} .
 \label{eq12}
\ee
This means that, if the cloaking device is designed with inversion symmetry (it should
not matter whether we look through the front or through the back), then we can consider
as a convenient working hypothesis that $d_1 = d_2 = L/2$, which leads to $f_1 = f_3 = f$.
Making the corresponding substitutions in either Eq.~(\ref{eq11a}) or Eq.~(\ref{eq11d}),
with the cloaking conditions (\ref{eq8a}) or (\ref{eq8d}), and solving for $f_2$, we obtain
a non-vanishing value for this focal length:
\be
 f_2 = \frac{L - 2f}{4} .
 \label{eq13}
\ee
It can readily be noticed that, if this condition is substituted into (\ref{eq11c}),
this matrix element vanishes.
The resulting matrix then reads as
\be
 \mathbb{M}_3 =
 \left( \begin{array}{cc}
  1 & L/(1 - L/2f) \\ 0 & 1
  \end{array} \right) ,
 \label{eq14}
\ee
which, for $2f \gg L$, can be recast as
\be
 \mathbb{M}_3 \approx
 \left( \begin{array}{cc}
  1 & L \left( 1 + L/2f \right) \\ 0 & 1
  \end{array} \right) .
 \label{eq15}
\ee
If the term $L/2f$ can be neglected ($L \ll f$), then the matrix (\ref{eq15}) acquires
the form of (\ref{eq7}) and, in principle, this condition might allow cloaking.
Moreover, when applied to Eq.~(\ref{eq13}), this assumption implies that $f_2$ must
be a negative lens, since $f_2 \approx - f/2$, unlike $f_1$ and $f_3$, which are both
positive (convergent).
Therefore, we find that the first non-trivial OCD can be achieved by playing with
lenses with relatively large focal lengths and different vergence.


\subsection{Four-lens system}
\label{sec25}

Let us now consider a setup consisting of four lenses.
Proceeding as before, the elements of the corresponding $\mathbb{M}_4$
matrix read as
\ba
\fl A_4 & = & 1 - \frac{L}{f_1} - \left( 1 - \frac{d_1}{f_1} \right)
     \left( d_2 + d_3 - \frac{d_2 d_3}{f_3} \right) \frac{1}{f_2}
   - \left[ 1 - \frac{(L - d_3)}{f_1} \right] \frac{d_3}{f_3} ,
 \label{eq16a}
 \\
\fl B_4 & = & L - \frac{\left( L - d_1 \right) d_1}{f_2}
   - \frac{\left( L - d_3 \right) d_3}{f_3} + \frac{d_1 d_2 d_3}{f_2 f_3} ,
 \label{eq16b}
 \\
\fl C_4 & = & - \frac{1}{f_1} - \frac{1}{f_2} - \frac{1}{f_3} - \frac{1}{f_4} + \frac{L}{f_1 f_4}
   + \left( \frac{1}{f_2} + \frac{1}{f_3} \right) \left( \frac{d_1}{f_1} + \frac{d_3}{f_4} \right)
 \nonumber \\
\fl & & + \left( \frac{1}{f_1 f_3} + \frac{1}{f_2 f_4} + \frac{1}{f_2 f_3} \right) d_2
   - \left[ \frac{\left( L - d_1 \right) d_1}{f_2} + \frac{\left( L - d_3 \right) d_3}{f_3} \right]
     \frac{1}{f_1 f_4}
 \nonumber \\
\fl & &  - \left( \frac{d_1}{f_1} + \frac{d_3}{f_4} - \frac{d_1 d_3}{f_1 f_4} \right) \frac{d_2}{f_2 f_3} ,
 \label{eq16c}
 \\
\fl D_4 & = & 1 - \frac{L}{f_4} - \left( 1 - \frac{d_3}{f_4} \right)
     \left( d_1 + d_2 - \frac{d_1 d_2}{f_2} \right) \frac{1}{f_3}
   - \left[ 1 - \frac{(L - d_1)}{f_4} \right] \frac{d_1}{f_2} ,
 \label{eq16d}
\ea
with $L = d_1 + d_2 + d_3$.
Again here we can notice a certain symmetry by exchange of indices (both in focal lengths
and in inter-lens distances) that can be advantageously considered in order to determine
optimal cloaking conditions.

Thus, as before, let us assume the OCD satisfies inversion symmetry.
This means that the focal lengths of the outermost lenses is equal ($f_1 = f_4 = f_\alpha$),
and the same holds for the innermost lenses ($f_2 = f_3 = f_\beta$).
Moreover, in order to preserve such symmetry, it is also required that the distances $d_1$
and $d_3$ are equal.
So, from now on, $d_1 = d_3 = d_\alpha$ and $d_2 = d_\beta$.
With this, the matrix elements specified by Eqs.~(\ref{eq16a}) to (\ref{eq16d}) can be recast as
\ba
\fl A_4 & = & 1 - \left( \frac{1}{f_\alpha} + \frac{1}{f_\beta} \right) L
   + \frac{2 d_\alpha (L - d_\alpha)}{f_\alpha f_\beta}
   + \frac{d_\alpha d_\beta}{f_\beta^2} - \frac{d_\alpha^2 d_\beta}{f_\alpha f_\beta^2} ,
 \label{eq17a}
 \\
\fl B_4 & = & L - \left( L - d_\alpha \right) \frac{2 d_\alpha}{f_\beta} + \frac{d_\alpha^2 d_\beta}{f_\beta^2} ,
 \label{eq17b}
 \\
\fl C_4 & = & - \frac{2}{f_\alpha} - \frac{2}{f_\beta} + \frac{L}{f_\alpha^2} + \frac{d_\beta}{f_\beta^2}
   + \frac{2 L}{f_\alpha f_\beta} - \frac{2 \left( L - d_\alpha \right) d_\alpha}{f_\alpha^2 f_\beta}
   - \frac{2 d_\alpha d_\beta}{f_\alpha f_\beta^2}
   + \frac{d_\alpha^2 d_\beta}{f_\alpha^2 f_\beta^2} ,
 \label{eq17c}
 \\
\fl D_4 & = & 1 - \left( \frac{1}{f_\alpha} + \frac{1}{f_\beta} \right) L
   + \frac{2 d_\alpha (L - d_\alpha)}{f_\alpha f_\beta}
   + \frac{d_\alpha d_\beta}{f_\beta^2} - \frac{d_\alpha^2 d_\beta}{f_\alpha f_\beta^2} .
 \label{eq17d}
\ea
From the application of (\ref{eq8b}) to (\ref{eq17b}), we obtain
\be
 f_\beta = \frac{d_\alpha d_\beta}{2(L - d_\alpha)} ,
 \label{eq18}
\ee
which avoids making further assumptions on the relative size of $f_\beta$ with respect to $L$,
as in the previous cases, and hence allows some freedom of choice.
With this result and applying (\ref{eq8a}) to (\ref{eq17a}), we find the value of other focal length,
\be
 f_\alpha = \frac{d_\alpha L}{2(L - d_\alpha)} .
 \label{eq19}
\ee
As it can easily be noticed, the addition of these two focal lengths satisfies the relation
\be
 d_\alpha = f_\alpha + f_\beta .
 \label{eq20}
\ee
This means that the configuration of the OCD is such that the set of lenses 1 and 2, on the
one hand, and the set of lenses 3 and 4, on the other hand, form each a telescope, one in
front of the other.
Within this configuration, lenses 1 and 4 play the role of the objective, while lenses 2 and 3
would be the eyepieces, since $f_\alpha > f_\beta$, as it is inferred from the relation
\be
 \frac{f_\alpha}{f_\beta} = \frac{L}{d_\beta} .
 \label{eq21}
\ee
This relation also gives the magnification of each telescope.
Thus, we can see that the magnified image of a given object allocated before the first telescope
is reversed by the other, so that the total magnification becomes unitary.
This is precisely the counterpart in geometrical optics of the invisibility recipe in terms
of transfer matrices found by S\'anchez-Soto and coworkers for general electromagnetic fields
\cite{sanchezsoto:EJP:2008,sanchezsoto:PhysRep:2012}.
Furthermore, it should also be noticed that the condition leading to the cancelation of the
element $C_4$, in compliance with the functional form (\ref{eq7}), is precisely (\ref{eq20}),
which can easily be shown by direct substitution.

Taking into account that $L = 2d_\alpha + d_\beta$ in (\ref{eq21}) and the value of $d_\alpha$,
given by (\ref{eq20}), we can now obtain the value for $d_\beta$, which reads as
\be
 d_\beta = \frac{2 \left( f_\alpha + f_\beta \right) f_\beta}{f_\alpha - f_\beta} .
 \label{eq22}
\ee
This provides us with an exact solution (condition) for optical cloaking if we have two
sets of two lenses with focal lengths $f_\alpha$ and $f_\beta$.
The total size of the cloaking device will be
\be
 L = 2 d_\alpha + d_\beta
   = \frac{2 \left( f_\alpha + f_\beta \right) f_\alpha}{f_\alpha - f_\beta} .
 \label{eq23}
\ee


\section{Experimental implementation}
\label{sec3}


\subsection{General aspects}
\label{sec31}

In the analysis presented in the previous section, cloaking has been investigated within an ideal scenario based on the
following assumptions:
\begin{itemize}
 \item Paraxial conditions are always guaranteed.
 \item Lenses are aberration-free.
 \item Lack of aperture effects associated with the diameter of the lenses considered.
\end{itemize}
Obviously, these are ideal conditions that simplify the theoretical analysis, as we have seen above, but that have to be taken
into account when considering sets of standard lenses, as it is the case here, where we actually were not so much concerned
about getting a high degree of cloaking as constructing a relatively simple and cheap device that would allow us to study this
phenomenon.
In any case, all these handicaps are advantageous from a teaching perspective, since they can be used to better characterize
the cloaking conditions.
Furthermore, this is the reason why a projective OCD has been chosen to the detriment of a more appealing direct-sight OCD.

The so-called projective OCD prepared here is based on some basic optical properties.
Consider we illuminate a transparent slide and project its image on a somehow distant wall (or projection screen if the optical
bench is long enough).
When the OCD is inserted between the object and its image, if it has been properly implemented, its presence should not
affect too much the image (or not, at least, to a great extent), except for a reduction of luminosity due to the many lenses
involved in the setup.
Moreover, if an additional object is inserted inside the OCD (the ``hidden object''), its presence should not either affect
the image projected on the wall.
In spite of the difference in its performance with respect to a direct-sight OCD, notice that the theory introduced in the
previous section is still applicable, since the operation principle is the same (i.e., it does not matter whether we look through
the OCD or we make the light from an object to cross it and form an image beyond it).

\begin{figure}[t]
 \centering
 \includegraphics[width=0.85\textwidth]{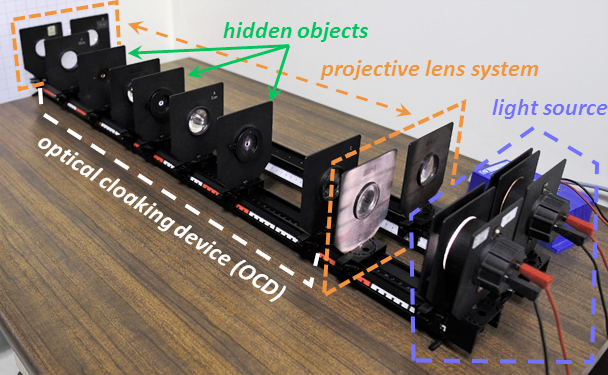}
 \caption{\label{Fig1}
  Full OCD experimental setup used here.
  In order to better appreciate the cloaking effect, a ``twin'' imaging system is accommodated side by side,
  so that the images produced by both are compared.
  As it can be seen, the setup consists of four basic elements: a light source with the object (a transparent
  slide with different shapes depicted), the projective lens system (see Sec.~\ref{sec32}), the OCD itself,
  and a series of diaphragms with variable diameter.}
\end{figure}

A photograph of the full experimental setup implemented here to investigate optical cloaking, as it has been used
during the different experiments carried out, is displayed in Fig.~\ref{Fig1}.
It consists of two standard optical benches with a length of about one meter and a half, a set of simple teaching lenses
with different focal lengths, several iris-type diaphragms, and two halogen lamps connected to 6~V/12~V DC power
suppliers.
The setup essentially consists of the object light sources (surrounded by a blue dashed line), the projective system
(orange dashed line), and the OCD (embraced by a white dashed line).
The hidden objects (plates with iris-type diaphragms) are all at display (their positions denoted with green arrows),
although depending on the experiment performed they could all be mounted, or only one of them.

We have considered two optical benches, because the OCD is going to be mounted in one of them, while the other
is just an idler partner.
The purpose of the idler bench is to produce a reference image that is not affected by any of the effects due to
either the OCD or the hidden object.
These two benches were aligned side-by-side very close together, such that the image produced with the OCD
could be easily compared with the idler one by direct sight (it was difficult, though, to get an optimal photograph
of them, as it can be inferred from Fig.~\ref{Fig4}).
Thus, when the OCD is not mounted, the two optical systems produce exactly the same image (see Sec.~\ref{sec32});
when the OCD is mounted, the presence of the idler image allows us to determine to what extent cloaking is achieved
and its quality (particularly, to detect any change in the size of the image, presence of aberrations or decrease in the
luminosity).

The lenses considered have focal lengths of $+5$~cm, $+10$~cm, and $+20$~cm, all of them with a diameter
of 3.5~cm and mounted on opaque $10\times 11$~cm$^2$ square frame plates.
These lenses are used to construct both the projective system and the OCD (see below).
As mentioned above, the role of hidden object is played here by a series of iris-type diaphragms mounted on plates with
the same features as those for the lenses.
These diaphragms have all of them maximum and minimum diameters of 3~cm and 0.1~cm, respectively.
They allow us to determine the maximum area that can be covered inside the OCD, while still observing the image
without much distortion (aberration effects) or remarkable loses of luminosity.
Or, in other words, the regions around the transferred bundle of rays where an object can be hidden without noticing
its presence.
It is worth stressing the fact that, because of the negligible thickness of the stop blades (and even the frame
where they are accommodated), this working method is ideal to determine the optimal distances or longitudinal ranges
(alongside the bench) where neither the position or the diameter of the diaphragm (or diaphragms) affect too much
the projected image, that is, the tolerance ranges of the OCD to hide object (see discussions in this respect
in Sec.~\ref{sec32} regarding the different experiments carried out and the corresponding tolerance ranges found).


\subsection{The projective imaging system}
\label{sec32}

\begin{figure}[t]
 \centering
 \includegraphics[width=\textwidth]{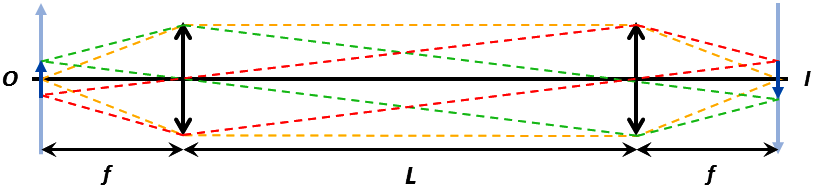}
 \caption{\label{Fig2}
  Ray-tracing diagram showing the effects on imaging due to size limitation of the diameter of the lenses
  used in the projective system.
  As it can be seen, due to such limitation, not every point of the illuminated object has an image.
  The original object is denoted with the shaded blue arrow on the left and its expected image with the inverted
  shaded blue arrow on the right.
  The final object and image are represented with the dark blue arrow (left and right, respectively).
  Limiting rays for the on-axis object point are displayed with orange dashed line, while for off-axis points are denoted
  green and read dashed lines (arising from the top and bottom parts of the object, respectively).}
\end{figure}

Previous to the practical implementation of the OCD, it is necessary to prepare the projective system that ensures a real
image with unitary linear magnification.
To that end, we have considered two lenses with the same focal length, $f = + 10$~cm.
One lens is positioned approximately at the focal distance away from the object (10~cm), while the other is at about the
same distance from the wall.
This particular arrangement, where both object and image are accommodated at the focal planes of the lenses
(front and rear, respectively), ensures the production of a real image with the same size as the object, although
inverted.
This can easily be seen from the ray diagram displayed in Fig.~\ref{Fig2}, by inspecting the green and red rays (dashed
lines) leaving, respectively, the top and bottom points of the object (dark blue arrow on the left).
Nevertheless, in terms of the transfer matrix method, if we construct the matrix from the object plane to the image one,
we have:
\ba
\fl
 \mathbb{M}_{IO} & = &
  \left( \begin{array}{cc} 1 & f \\ 0 & 1 \end{array} \right)
  \left( \begin{array}{cc} 1 & 0 \\ - 1/f & 1 \end{array} \right)
  \left( \begin{array}{cc} 1 & L \\ 0 & 1 \end{array} \right)
  \left( \begin{array}{cc} 1 & 0 \\ - 1/f & 1 \end{array} \right)
  \left( \begin{array}{cc} 1 & f \\ 0 & 1 \end{array} \right) \nonumber \\
\fl
 & = & \left( \begin{array}{cc} - 1 & 0 \\ - 2/f + L/f^2 & - 1 \end{array} \right) ,
 \label{eqIO}
\ea
which is a symmetric matrix (it displays the same functional form going from the object $O$ to the image $I$,
as in the other way around).
As it can be noticed, the lateral magnification (element $A$) and the angular one (element $D$) are both equal
to $-1$, which denotes the fact that the image is inverted with respect to the object, although the size is the
same.
Regarding the element $C$, it corresponds to the equivalent power for two identical lenses separated a distance
$L$, according to Gullstrand's equation \cite{milton-bk}, while a vanishing element $B$ denotes the fact that
the input plane corresponds to the front (object) focal plane of the system and the output plane to the rear
(image) focal plane.

In Fig.~\ref{Fig2} it can also be noticed that the full object $O$, denoted with the shaded blue arrow on the left,
is not going to produce an image $I$ (shaded blue arrow on the right).
This arises as a consequence of the above commented effect of the size-limitation of the lenses, which prevents rays
coming from any point on the object to pass through the two-lens system and produce a full image.
Typically, ray tracing in paraxial optics assumes that extension of the object is relatively small compared to the diameter
of the lens, by virtue of which sine and tangent functions can be approximated by the value of their arguments.
In realistic optical systems, like the one we are dealing with here, where the object is a rectangular transparent slide
of several centimeters wide and high, while the diameter of the lenses is smaller, the approximation works fine only for
object points off the system optical axis but still close to it; as object points become more and more off the optical
axis, the approximation breaks and additional considerations are required in order to explain or determine the imaging
process.
In principle, this leads to introduce some more advanced technical knowledge on aperture and field stops.
However, in order to keep the discussion here at the simplest level, which is one of the main purposes of the work,
we are going to further exploit the diagram of Fig.~\ref{Fig2} and extract such an information directly from it.

Thus, consider again the object denoted with the left shaded blue arrow.
Any bundle of rays leaving any point along this object will be able to pass through the front lens.
If the object is accommodated on the front focal plane of the lens, then ray bundles leaving the same point will
emerge parallel from the lens.
In the case of the on-axis object point, this is illustrated by the two orange dashed lines.
As it can be seen, after reaching the rear lens, the corresponding bundle of parallel rays will merge into the on-axis
image point, at the back focal plane of such a lens.
Now, if such ray bundles are also required to pass through a second lens, namely the rear lens here, this constitutes
a severe restriction, because not all the parallel ray bundles leaving the front lens will be able to reach totally or even
partially the rear lens.
There will be ray bundles with such an inclination that, after having travel the distance $L$, will fall out of the diameter
of the rear lens.
For example, in the diagram of Fig.~\ref{Fig2} we notice that, compared to the on-axis object point, only a half of the
ray bundles leaving the front lens can reach the rear one if such rays come either from the top of the dark blue arrow
(see green dashed lines) or the bottom (red dashed lines).
When these bundles cross the rear lens, the merge respectively into the bottom and top off-axis image points, denoted
with the dark blue arrow on the right.
It is clear that, because only half of the initial bundle that penetrated the front lens is going to reach the focal plane of
the rear lens, the luminosity of the image in those points will be lesser than closer to the optical axis.
The same can be applied to object points further away from the optical axis, with a relatively quick loss of luminosity in
the corresponding image points.

In our particular case, taking also into account the points for which the ray bundles reaching the rear lens reduce to
a half, and considering that the lens diameter of 3.5~cm and a distance between both lenses $L = 84.7$~cm, a simple
trigonometry-based calculation renders an estimate for the size of the image of about 0.41~cm.
This is the same to say that only an effective circular spot in the object with such diameter is going to form a clear image,
even if the illuminated area of the object is much larger.
Nonetheless, in practice we have noted that the image is a bit larger, namely a spot of about 0.7~cm (see Fig.~\ref{Fig5}),
which means that ray bundles coming from upper or lower points in the object are also going to contribute, although with
smaller luminosity and importantly affected by spherical and chromatic aberrations.
For instance, if we consider the limit of the off-axis object points for which only one of the corresponding outgoing rays is
going to pass through both lenses, we find an estimate of the spot diameter of about 0.83~cm, although the borders of
such a spot will be relatively dark.
By averaging with the previous value, we obtain a spot size of 0.62~cm, which is closer to the value observed experimentally.
This means that even object points contributing with about less than a quarter of the ray bundle that passes through the front
lens are going to be significant.
In any case, these values are fine, because what we have used as a test object/image is a picture of two parallel straight
segments of 0.2~cm length.
Notice that with a smaller OCD, $L$ would also be smaller and therefore the projected image would be larger.

Regarding this projective system, it is also worth mentioning that, if the light source and the two projective lenses are removed,
and in the place of the original image on the wall we put a picture, when looking through the OCD we can see (although
affected by some amount of aberration) the image of a such picture with exactly the same size and orientation.
This experiment was performed in order to confirm the conditions (i) and (ii) by direct sight, although the result was not
as spectacular as in Ref.~\cite{choi:OptExp:2014} and it was not possible to obtain any good quality photograph to be
reported here.
Furthermore, also notice that analogous size-limitation effects are going to be associated with the lenses of the OCD itself,
although in a minor proportion, as it will be seen.


\subsection{Experiments with the projective OCD}
\label{sec33}

\begin{figure}[t]
 \centering
 \includegraphics[width=\textwidth]{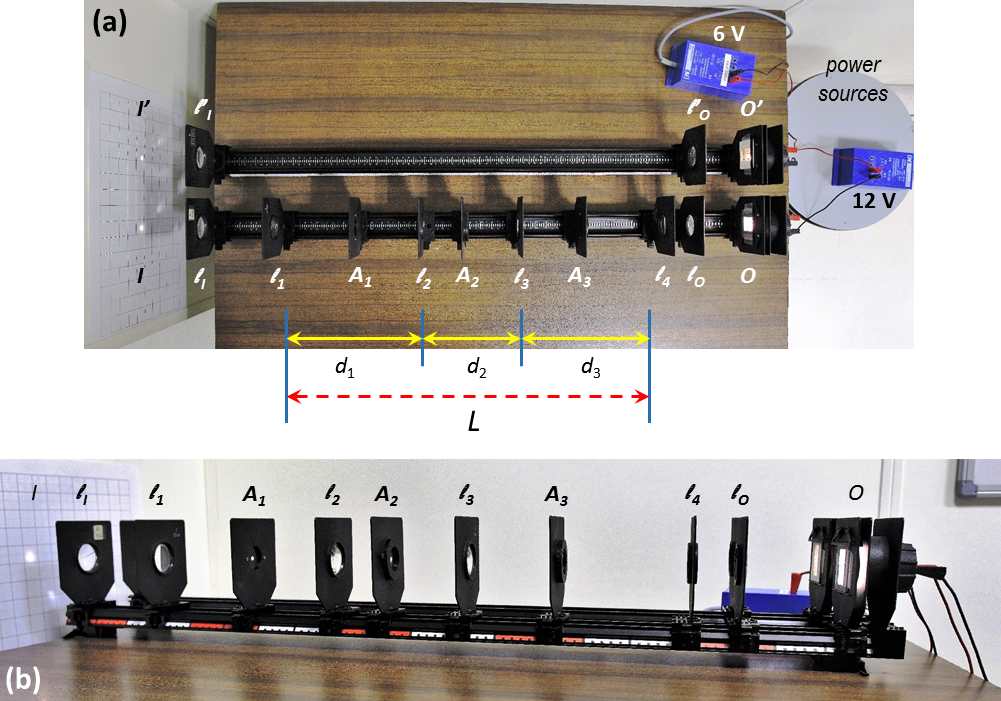}
 \caption{\label{Fig3}
  Top view (a) and side view (b) of experimental setup used as OCD, where $\ell_i$ ($i = 1, 2, 3, 4$) denote
  the positions of the lenses used and $A_j$ denotes the positions of three ($j = 1, 2, 3$) iris-type diaphragms.
  The distances between lenses $d_k$ (with $k = 1, 2, 3$) and the total length $L$ of the device are also shown.
  Notice in panel (a) that both projective systems, test and OCD, are replicas one another.}
\end{figure}

Figure~\ref{Fig3} shows a top view (a) and a side view (b) of the full experimental setup
used here.
The top view in panel (a) shows how the OCD almost occupies the full length between the lenses
of the projective system (compare the lower bench, where the OCD is mounted, with the idler,
which is upper one), as well as the proximity between both benches.
The side view, in panel (b), gives an idea of the side-by-side alignment, very beneficial for
a direct comparison of the images $I$ (with the OCD) and $I'$ (idler).

The OCD implemented can be better seen in Fig.~\ref{Fig3}, where a top view (a) and a side view (b) are shown.
For an easier identification of the different elements, the projective system lenses are denoted with $\ell_O$ for
the front lens and $\ell_I$ for the rear one (the same for the idler companion, but with primes).
The lenses constituting the OCD are denoted as $\ell_i$, with $i = 1, 2, 3, 4$.
This criterion follows the same labeling used in the theoretical section, where the closest lens to the observer
is precisely the one closer to image in our case here.
Accordingly, we have also labeled the three spaces generated by every two consecutive lenses as 1, 2 and 3,
increasing from the image to the object.
Hence, the distances spanned by these spaces are $d_1$, $d_2$ and $d_3$, with the total length of the OCD being
$L = d_1 + d_2 + d_3$.
As for the diaphragms, to keep a consistent notation, they are referred to as $A_1$, $A_2$ and $A_3$ ($A$ for
aperture), which makes reference to the space where they are accommodated, although they all are identical.
Because of the decrease of luminosity in the image when the OCD is introduced, in the corresponding setup
a power of 12~V is supplied to the halogen bulb, although only 6~V were required in the idler companion,
otherwise the image spot was too bright under darkness conditions, which were used for a better performance
of the experiment (the photos of Figs.~\ref{Fig1} and \ref{Fig3} were taken with daylight conditions).

\begin{figure}[t]
 \centering
 \includegraphics[width=\textwidth]{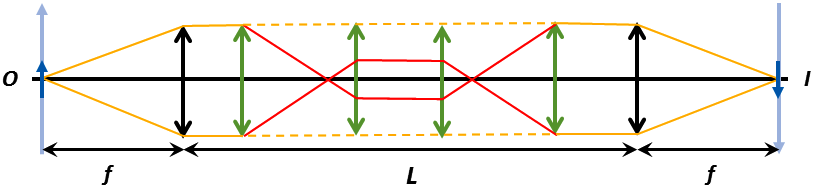}
 \caption{\label{Fig4}
  Ray diagram illustrating the working principle of the OCD implemented here: the rays
  going from the object to the image (orange solid lines) pass through the device as if
  it was not there (orange dashed lines) by deviating them (red solid lines).}
\end{figure}

Figure~\ref{Fig4} illustrates by means of a simple ray diagram the working principle of the OCD:
the rays leaving the on-axis object point (orange solid lines) pass through the device as if it
was not there (dashed lines).
This transit takes place by diverting the rays incident onto the OCD in the way indicated by
the red solid lines.
Notice that this ray diversion is theoretically described by the matrix found in Sec.~\ref{sec25}, which would
replay the central transit matrix in Eq.~(\ref{eqIO}), in compliance with the fact that the effect of the four-lens
transfer matrix should be equivalent to having nothing along the path $L$ pursued by the incoming object rays.
Specifically, the lenses selected to built the OCD have focal lengths of $+20$~cm (for $\ell_1$ and $\ell_4$)
and $+5$~cm ($\ell_2$ and $\ell_3$).
Accordingly, from Eq.~(\ref{eq20}), the distance between them is $d_1 = d_3 = 25$~cm, while
$d_2 \approx 16.7$~cm, from Eq.~(\ref{eq22}).
The OCD was mounted in such a way that $\ell_4$ was at 5.4~cm from $\ell_O$ and $\ell_1$ at 12.6~cm from $\ell_I$.
As mentioned above, cloaking conditions have been investigated by using iris-type diaphragms.
It is clear that, as the diaphragm diameter is decreased, the bundle of rays will also decrease, which have
an observable effect on the image.
Therefore, by conveniently choosing a relatively narrow aperture that still allows the full bundle
of rays to pass through, it is possible to determine a range of positions of the top along which its
presence will not alter the image.
In other words, the presence of the diaphragm will be cloaked unless its diameter is so small that it starts
affecting the projected image.
Several experiments were carried out with analogous results.

\begin{figure}[t]
 \centering
 \includegraphics[width=0.95\textwidth]{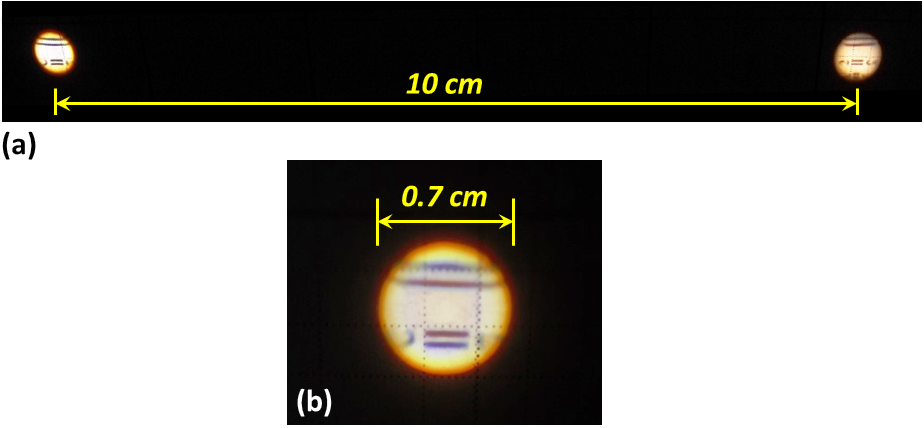}
 \caption{\label{Fig5}
  (a) Photograph of the illuminated spots that can be observed during the performance of
  the experiments reported here.
  The left spot corresponds to the bench with the OCD, while the right spot is the idler one.
  Although they are circular, here the present an oval shaped due to the perspective introduced
  by the camera.
  (b) Enlargement of the left spot showing the colored halo produced by chromatic aberration
  in the limit of the optimal cloaking region (it appears much more enhanced that it is
  actually, because the photograph has received a high-contrast treatment in order to better
  visualize it).}
\end{figure}

The first experiment performed consisted in determining the optimal cloaking conditions
for each diaphragm individually considered in the setup.
Notice that, if the diaphragm diameter is reduced in an important amount, by moving it
along the corresponding section of the OCD we should obtain a spatial range for which
the projected image is not affected.
This range will be the optimal cloaking region, that is, an observer at the position of
the projected image will not be able to perceive the presence of any object accommodated
within such a range beyond the boundary defined by the corresponding iris diameter.
With this in mind, we selected an aperture of 0.5~cm for all three diaphragms, which
corresponds to about a 2.8\% of their maximum area when they are fully open.
Accordingly, we have found the following optimal distances (tolerance ranges):
\begin{itemize}
 \item For $A_1$, with a range going from 11.7~cm to 13.2~cm measured from $L_1$, no important
 effects were observed in the projected image, such as loss of luminosity or appearance of
 chromatic aberrations (in the form of light color rings surrounding the image).
 This range lies around the center of the section (at 12.5~cm from $L_1$).

 \item For $S_2$, the range goes from 6.7~cm to 8.9~cm, measured from $L_2$, which is also
 around the center of the section ($\approx 8.4$~cm from $L2$).

 \item For $S_3$, the range was between 9.7~cm and 11.6~cm, measured from $L_3$, closer
 (although still below) the center of the section (12.5~cm from $L_3$).
\end{itemize}
A photograph of what can be seen projected onto the wall during the performance of
the experiment is displayed in Fig.~\ref{Fig5}.
The two images, the one produced with the OCD (left) and the idler one (right), are shown
together for comparison in panel (a).
Due to the small size of the illuminated spots, the distance between them seem to be
relatively large, although there are only 10~cm between their centers.
Actually, the distortion observed (the kind of oval shape that they display) is due to
the perspective introduced by the camera.
In panel (b) we show only the image produce with OCD, where the colored halo due to the
incipient effects of the chromatic aberration can be seen.
Nevertheless, it is worth mentioning that the effect is much stronger than it is
actually, because of the treatment of the image, which require a high contrast for
a better visualization.

Once cloaking ranges are set on each section of the OCD, one might wonder about the maximum
cloaking achievable with this device.
By inspecting Fig.~\ref{Fig4}, given that the ray bundle crossing the OCD is going to get
narrower between $\ell_2$ and $\ell_3$, we are going to focus on this section.
By moving the diaphragm $A_2$ from one of these lenses to the other and making narrower its
diameter, we finally found such optimal cloaking conditions at 7.7~cm from $L_2$ (near the
central value found in the previous experiment), for a diameter of 0.2~cm, i.e., the diaphragm
opening is only about 0.4\% of its maxima opening.
For these conditions, we noticed that:
\begin{itemize}
 \item If the diameter was decreased at this position, there was a remarkable quick
 loss of luminosity in the projected image.

 \item If the diaphragm was slightly displaced backwards, towards $L_2$, vignetting-related
 effects became remarkable.

 \item If the diaphragm was slightly displaced forward, towards $L_3$, the effects of chromatic
 aberrations also appeared immediately (blue spots inside the image).
\end{itemize}

One might also wonder why instead of considering the central section of the OCD, the first or
third sections were not considered in the regions where the rays cross the foci of its lenses.
The reason is very simple: although there is a minimum (zero) waste there for incident rays
that are parallel to the optical axis, rays incident with any other inclination will not
pass through such points, which constitutes an important inconvenience.
Nevertheless, this led us to consider a third experiment to test the robustness of the cloaking
condition rendered by the previous experiment.
That is, without changing either the diameter or the position of the diaphragm $A_2$, we decided
to determine where inside the other sections of the OCD an object could be hidden without noticing
its presence.
Thus, we selected a diameter of 0.4~cm for the other two diaphragms, namely $A_1$ and $A_3$.
With this diameter we ensured a reduced loss of luminosity in the projected image when all three
diaphragms were used.
The optimal positions for these diaphragms were 12.1~cm for $A_1$, measured from $\ell_1$, and
10.7~cm for $A_3$, measured from $\ell_3$.
Notice that none of them is close to the corresponding focus, but they are closer to the center
of the corresponding section of the OCD.
Furthermore, also notice that the two positions lie close to the respective central values of
the ranges found in the first experiment.


\section{Concluding remarks}
\label{sec4}

The main purpose of this work has been to provide startup tools to introduce the
optical cloaking phenomenon in the classroom both at the theory level and also at
the level of teaching experiment.
This has been done by exploring the performance of what we have called a projective
optical cloaking device (OCD), where instead of directly observing through the device
itself, as it is the case in Ref.~\cite{choi:OptExp:2014}, the phenomenon is studied
and analyzed by inspecting the projected image of an illuminated object.
The main advantage of this setup is that it can be built with material that is currently
available in any optics teaching laboratory (organic teaching lenses), without the
necessity to rely on high-quality material, such as good quality lenses (glass research
lenses).
By means of a series of experiments (these are just some examples, but many others
can be devised) has served to detect the phenomenon and also to determine optimal
cloaking regimes for the setup considered.
In this regard, we have advantageously used the fact that the lenses were not
aberration-free, which has allowed us establishing appropriate boundaries for the
cloaking regimes.
These regimes have been found to happen in all three sections of the OCD, with
cloaking ranges of about 2~cm, approximately, within each section, although in
principle one would expect to detect cloaking only along the central section, where
the incident ray bundle gets narrower, as it is illustrated in Fig.~\ref{Fig3}.
In this sense, we have also seen how putting a number of lenses one after the other
becomes very important regarding cloaking, not only because of a remarkable reduction of
luminosity, but also because of aperture issues, exemplified by means of Fig.~\ref{Fig2}.

The OCD here has been built taking into account a theoretical analysis based on the transfer
matrix formulation of Gaussian paraxial optics.
The main reason why we have chosen this method instead of more conventional ones
based on ray tracing is because it stresses in a nice manner the input/ouput relationship
enabled by the system analyzed.
Although the transfer matrix method is not widely known, it is worth introducing at this
level, because the construction of the system matrix allows to get a general view of
the path followed by the rays, since they depart from the object until they reach the
projection wall where the image is formed, without restricting ourselves to a limited set
of rays.
Nevertheless, as it can be seen, it is not necessary a high knowledge on the method,
but just a few aspects; the rest is just standard matrix algebra.
On the other hand, the appealing feature of ray tracing constructions, however, is not lost
either, because they have also been used here, in particular to illustrate the passage of rays
through the projective system or the OCD, as shown respectively in Figs.~\ref{Fig2} and
\ref{Fig3}.
In general terms, we have found that, in the same way that the experimental OCD is
simple, the theory here considered has also be presented at a simple level in order to
make it suitable for undergraduate optics courses.
In this regard, the introduction of more sophisticated, on-purpose software typically used
in this kind of analysis either to proceed with the ray tracing or to compute and solve matrices
has been skipped, because the main idea is to tackle the problem just with simple tools.
For the same reason, a more refined and explicit analysis of the problem based on the role
of lenses as aperture and field stops has also been omitted, because it already requires
some more advanced knowledge on the issue, which are not really necessary at this stage
to explain imaging, as we have shown.


\ack

Financial support from the Spanish MINECO (Grant No.\ FIS2016-76110-P)
is acknowledged.


\section*{References}


\begin{thebibliography}{10}
\expandafter\ifx\csname url\endcsname\relax
  \def\url#1{{\tt #1}}\fi
\expandafter\ifx\csname urlprefix\endcsname\relax\def\urlprefix{URL }\fi
\providecommand{\eprint}[2][]{\url{#2}}

\bibitem{rowling:HarryPotter:2007}
Rowling J~K 2007 {\em Harry Potter and the Deathly Hallows\/} 1st ed (New York:
  Bloomsbury Publishing)

\bibitem{steinmeyer-bk}
Steinmeyer J 2003 {\em Hiding the Elephant -- How Magicians Invented the
  Impossible and Learned to Disappear\/} (New York: Carroll \& Graf Publishers)

\bibitem{zepf:physteach:2004}
 Experiments from \cite{steinmeyer-bk} were used at Creighton University for a
 sesion of recreational physics:
 Zepf T~H 2004 {\em Physics Teacher\/} {\bf 42} 404--408

\bibitem{azanna:optica:2018}
Cort\'es L~R, Seghilani M, Maram R and Aza{\~n}a J 2018 {\em Optica\/} {\bf 5}
  779--786

\bibitem{engheta:PRE:2005}
Al\`u A and Engheta N 2005 {\em Phys. Rev. E\/} {\bf 72} 016623(1--9)

\bibitem{engheta:PRE:2006}
Al\`u A and Engheta N 2006 {\em Phys. Rev. E\/} {\bf 73} 019906(E)

\bibitem{smith:Science:2006}
Schurig D, Mock J~J, Justice B~J, Cummer S~A, Pendry J~B, Starr A~F and Smith
  D~R 2006 {\em Science\/} {\bf 314} 977--980

\bibitem{shalaev:NatPhot:2007}
Cai W, Chettiar U~K, Kildishev A~V and Shalaev V~M 2007 {\em Nat. Photon.\/}
  {\bf 1} 224--227

\bibitem{marques:AJP:2011}
Vel\'azquez-Ahumada M~C, Freire M~J, Algar{\'\i}n J~M and Marques R 2011 {\em
  Am. J. Phys.\/} {\bf 79} 349--352

\bibitem{fleming:AJP:2017}
Fleming S 2017 {\em Am. J. Phys.\/} {\bf 85} 173--177

\bibitem{gennaro:AJP:2016}
Wilkinson J~T, Whitehouse C~B, Oulton R~F and Gennaro S~D 2016 {\em Am. J.
  Phys.\/} {\bf 84} 14--20

\bibitem{thompson:AJP:2008}
Sipos M and Thompson B~G 2008 {\em Am. J. Phys.\/} {\bf 76} 464--469

\bibitem{longhi:AJP:2017}
Horsley S~A~R and Longhi S 2017 {\em Am. J. Phys.\/} {\bf 85} 439--446

\bibitem{choi:ApplOpt:2014}
Howell J~C, Howell J~B and Choi J~S 2014 {\em Appl. Opt.\/} {\bf 53} 1958--1963

\bibitem{howell:video}
 A visual demostration of the work reported in \cite{choi:ApplOpt:2014} is available at:\\
 \url{http://www.rochester.edu/news/show.php?id=6522}

\bibitem{choi:OptExp:2014}
Choi J~S and Howell J~C 2014 {\em Opt. Exp.\/} {\bf 22} 29465--29478

\bibitem{choi:OptExp:2015}
Choi J~S and Howell J~C 2015 {\em Opt. Exp.\/} {\bf 23} 15857--15862

\bibitem{howell:video2}
 A visual demostration of the work reported in \cite{choi:OptExp:2014} is available at:\\
 \url{https://www.youtube.com/watch?v=vtKBzwKfP8E}

\bibitem{orbita-laika}
 A visual demostration of a former OCD version constructed in our teaching lab and shown
 recently in a the popular science TV show {\it Orbita Laika} (the demo starts around minute
 2'30'') is available at:\\
 \url{https://www.youtube.com/watch?v=gMV_Gm4QJWo}

\bibitem{pedrotti-bk}
Pedrotti F~L, Pedrotti L~M and Pedrotti L~S 2014 {\em Introduction to Optics\/}
  (UK: Pearson)

\bibitem{milton-bk}
Katz M 2002 {\em Introduction to Geometrical Optics\/} (New Jersey: World
  Scientific)

\bibitem{sanchezsoto:EJP:2008}
Monz\'on J~J, Barriuso A~G and S\'anchez-Soto L~L 2008 {\em Eur. J. Phys.\/}
  {\bf 29} 431--437

\bibitem{sanchezsoto:PhysRep:2012}
S\'anchez-Soto L~L, Monz\'on J~J, Barriuso A~G and Cari{\~n}ena J~F 2013 {\em
  Phys. Rep.\/} {\bf 513} 191--227

\end{thebibliography}

\providecommand{\newblock}{}

\end{document}